 \documentclass[showpacs,prb,twocolumn,superscriptaddress,floatfix,amsmat]{revtex4}
  \usepackage{graphicx,amsmath,color}
 
\begin{document}
 \title{Two-Component Scaling near the Metal-Insulator Bifurcation in Two-Dimensions}
 \author{D.J.W. Geldart}
 \affiliation{Department of Physics, Dalhousie University, Halifax, NS B3H3J5, Canada }
 \affiliation{School of Physics, University of New South Wales, Sydney 2052, Australia}
\affiliation{Laboratoire des Verres, Universit\'e de Montpellier
II, Montpellier 34095, France }
 \author{D. Neilson}
  \affiliation{School of Physics, University of New South Wales, Sydney 2052, Australia}
 \affiliation{Dipartimento di Fisica, Universit\`a di Camerino, 62032 Camerino, Italy}
 \begin{abstract}
 We consider a two-component scaling picture for the
resistivity of two-dimensional (2D) weakly disordered interacting electron
systems at low temperature with the aim of describing both the vicinity of the
bifurcation and the low resistance metallic regime in the same framework. We
contrast the essential features of one-component and two-component scaling
theories.   We discuss why the conventional lowest order renormalization group
equations do not show a bifurcation in 2D, and a semi-empirical extension is
proposed which does lead to bifurcation. Parameters, including the product
$z\nu$, are determined by least squares fitting to experimental data. An
excellent description is obtained for the temperature and density dependence of
the resistance of silicon close to the separatrix.  Implications of this
two-component scaling picture for a quantum critical point are discussed.
 \end{abstract}
  \maketitle

\section{Introduction}

The determination of the low temperature properties, and ultimately the ground
state phase diagram, of two-dimensional (2D) systems of strongly interacting
electrons at low carrier density in the presence of low levels of disorder
remains a very challenging problem in spite of a great deal of experimental and
theoretical effort. The longstanding view that all charge carrier states become
localized at zero temperature ($T=0$) in the limit of a large 2D system has
been called into question by  experimental observations of a finite temperature
"transition" from insulator-like ($\partial\rho/\partial T < 0$) to metal-like
($\partial\rho/\partial T > 0$) behavior  as the electron density is increased
in very high purity semiconductor MOSFETs and heterostructures
\cite{Kravchenko,Simonian,Coleridge,Simmons}. This bifurcation of the
resistivity $\rho(T)$ into two families of curves occurs at a critical carrier
density of the 2D electron (or hole) system. The critical carrier density
depends on the specific disorder characteristics of the given sample. In the
vicinity of the bifurcation, $\rho(T)$ has intriguing scaling properties as a
function of density and temperature. The lower density insulating family of
curves having $d\rho/dT<0$ can be collapsed onto a single curve when plotted as
a function of a scaling variable $T_0(\delta)/T$. The new density dependent
temperature scale $T_0(\delta)$ has  a power law dependence on the magnitude of
$\delta =(n-n^\star)/n^\star$ where $n^\star$ is the critical density at the
bifurcation. Similarly, the metallic family of curves at higher density with
$d\rho/dT>0$ collapses onto another unique curve with a power law dependence on
$T_0(\delta)/T$ with precisely the same critical exponent. There is also
scaling behavior with respect to density via $|\delta|$ and applied electric
field, with a different characteristic critical exponent.  A review has been
given by Abrahams {\it et al}\, \cite{RMP}.

A bifurcation with scaling behavior at finite $T$ is expected to be a generic
feature associated with a $T=0$ quantum critical point (QCP)\cite{Sondhi}.
Consequently, the demonstrated scaling for the resistivity can be taken as
evidence for a metal-insulator transition (MIT). With this assumption,
Dobrosavljevi\'c {\it et al} have given a  phenomenological description of the
observed scaling properties and power laws in terms of a one-parameter scaling
theory \cite{DAMC}. In this picture, the bifurcation point lies on a separatrix
which terminates at $T=0$ at a QCP which is a fixed point of a set of scaling
equations. The fixed point is repulsive with respect to the temperature
variable and separates the insulating and metallic regions of the low T phase
diagram. There is a unique correspondence between the QCP and the bifurcation
point and a two-phase ground state is implied.

However, this correspondence provides only indirect evidence for a MIT. Of
course the power law fits near the bifurcation point are based on linearization
of a scaling flow equation and the linearization must eventually fail at
sufficiently low temperature, for which $T_0(\delta)/T$ is no longer small, but
this is not evidence against a QCP. A more crucial point is that a physical
effect which is negligible for temperatures $T\sim 1$ K near the bifurcation,
may become dominant in the low $T$ limit.  Scaling properties characteristic of
an imminent QCP would then still be observed near the bifurcation but the
suggested QCP itself could be totally removed in the zero temperature limit.

A possible example of this situation is the proposal by Simmons
{\it et al}\, \cite{SimmonsWL} that in $p$-type GaAs the $\rho(T)$
curves of the metallic family will eventually turn upward to
exhibit insulator-like behavior with $\partial \rho/\partial T <
0$ if the electron temperature $T$ could be made sufficiently low.
A similar conclusion has recently been reached for Si MOSFETs.
\cite{Prus} However the turn-up is actually only directly
observed for metallic curves just above the critical density, and,
as we discuss below, the magnetoresistance data cited as evidence
of a universal weak localization in 2D in Refs.\
\onlinecite{SimmonsWL} and \onlinecite{Prus} may alternatively
reflect different magnetic field dependencies of the competing
localizing and delocalizing triplet spin state interactions.
Irrespective of the mechanism responsible for the upturn,
properties of the system at temperatures well below the
bifurcation temperature are not correctly described by a
one-component scaling theory.  We will outline a semi-empirical
two-component scaling theory that describes within a single
framework the temperature dependent resistivity both in the
vicinity of the bifurcation and in the low resistance metallic
regime.

Following early work by Altshuler {\it et al}\, \cite{AAL} further progress
toward an understanding of disordered 2D electron systems in the strong
coupling regime was made by Finkelstein \cite{Finkelstein1,Finkelstein2} and by
Castellani {\it et al}\, \cite{Castellani} who showed that
electron-electron interactions can lead to $d\rho/dT > 0$
(metallic behavior) in the disordered 2D system at finite $T$.
This is in contrast to the localizing insulating behavior found in
the absence of electron-electron interactions \cite{Abrahams}. The
physical properties of the system at low $T$ are determined by
nonlinear interactions of low energy diffusive modes. A set of
renormalization group (RG) equations was generated by means of a
formal perturbation expansion in powers of a dimensionless
temperature dependent resistivity ${\cal{R}}$ which is related to
the physical resistance per square by ${\cal{R}} = (e^2/\pi
h)R_{\fbox{\!}}$. These scaling results include electron-electron
interactions to all orders and are valid in the diffusive regime
$k_BT<\hbar/\tau$ provided ${\cal{R}}$ is small (in principle,
${\cal{R}}<<1$). Additional progress was made by Zala {\it et
al}\, \cite{Zala} who considered interaction corrections to
transport properties at intermediate temperatures in the ballistic
regime $k_BT>\hbar/\tau$ to all orders in the electron-electron
interaction using a Fermi-liquid approach. These results are also
restricted to the perturbative regime, ${\cal{R}}<<1$. Comparisons
with experiment for the temperature dependent resistivity of Si
MOSFETs have been made for both the scaling theory \cite{Punn} and
the intermediate temperature theory
\cite{Vitkalov,Pudalovinteraction}.  The intermediate temperature
theory gives consistent semi-quantitative agreement with
experiment in the range of densities for which ${\cal{R}}$ is
small.

We emphasize that both these theories are perturbative in ${\cal{R}}$ and are
limited to the regime of small ${\cal{R}}$.  Hence they must fail to describe
the regime near bifurcation where ${\cal{R}}\sim 1$ and they contain no
bifurcation point. In addition, in the RG procedure the equations develop
singularities at a nonzero  temperature so that the approach to the ground
state cannot be described.

In contrast to the 2D case there is a qualitative understanding of the
metal-insulator transition in 3D electron systems. A consistent qualitative
picture of the transition has been obtained by RG methods in a space of
dimension $2+\epsilon$ with $\epsilon>0$ taken as an expansion parameter.
Physical quantities such as critical exponents are expressed as power series in
$\epsilon$.  While only the leading terms in $\epsilon$ have been obtained it
is presumed that resummation methods would give good results in 3D if
sufficient correction terms could be calculated.  In the case of $\epsilon>0$
it is clear that the interplay of strong electron-electron interactions with
disorder is an essential aspect of the bifurcation.

It is remarkable that in spite of great effort a corresponding theory does not
exist for 2D systems, that is for $\epsilon=0$.  This lack casts doubt on the
physical significance  and relevance of the ``universal'' aspects of
electron-electron interactions in disordered systems. In addition the lack of
even an approximate scaling theory has prevented quantitative discussion of the
low temperature limit and the approach to the ground state.   We expect that
the interplay of strong electron-electron interactions with disorder must be an
essential aspect of the bifurcation in 2D irrespective of whether or not a true
metal-insulator transition occurs.  Of course in some  systems material
dependent effects also contribute.

In this paper we discuss why the lowest order RG equations  in 2D do not
show a bifurcation  and we give a semi-empirical extension which does describe
a bifurcation region. We focus on the immediate vicinity of the bifurcation and
on the metallic regime.  The strong insulator limit will be
discussed elsewhere.  The  proposed scaling equations have a physical
low temperature limit and have no singularities at finite length or temperature
scales.   This scaling picture provides a theoretical framework for the
interpretation of experimental results, more specifically the ``universal''
contributions due to electron-electron interactions and  disorder.

\section{Renormalization Group Equations}

The RG equations established at one-loop level for a 2D system of
electrons in the presence of disorder are based on four
dimensionless scaling parameters ${\cal{R}}$, $\gamma_2$, $Z$ and
$\gamma_c$, \cite{Finkelstein1,Finkelstein2,Castellani}  where
$\Gamma_2=Z\gamma_2$ is the electron-hole scattering amplitude for
the triplet spin state, $\Gamma_c=Z\gamma_c$ is the singlet state
particle-particle scattering amplitude and $Z$ is the dynamical
energy rescaling function. Together with ${\cal{R}}$, these
quantities are all functions of the variable $y=\ln\lambda^{-1}$
which describes rescaling of the momenta after integrating over
the momentum shell specified by $\lambda k_0^2<k^2<k_0^2$.
\cite{Castellani2}  We consider only the case when the disorder is
due to purely potential scattering.
When a connection between length and temperature scaling is
needed, we relate a thermal length $\ell_{th}$ to temperature by
$\ell_{th}/\ell_{el}=(T_{el}/T)^{1/z}$, where $\ell_{el}$ and
$T_{el}$ are the length and temperature scales for elastic
scattering and $z$ is the dynamical critical exponent
\cite{Sondhi}.

Particle-particle scattering is not considered to play an
essential role relative  to the electron-hole scattering
represented by $\gamma_2$ so we omit consideration of $\gamma_c$.
The energy scaling function $Z(y$) is not followed explicitly but
its effect is taken into account at the bifurcation fixed point by
$y=\ln[(T_{el}/T)^{1/z}]$. These approximations permit discussion
of a two-component scaling theory based on the scaling parameters
${\cal{R}}$ and $\gamma_2$.
With these simplifications the RG equations are
 \begin{eqnarray}
d{\cal{R}}/dy&=&\alpha(\gamma_2){\cal{R}}^2\\
 d\gamma_2/dy&=&\frac{(1+\gamma_2)^2}{2} {\cal{R}}
  \label{RGEqns}
  \end{eqnarray}
 where
 \begin{equation}
 \alpha(\gamma_2)=1+\left[n_v+((2n_v)^2-1)\left\{1\!-\frac{1\!+\gamma_2}{\gamma_2}
 \ln(1\!+\gamma_2)\right\}\right] \ .
\label{alpha}
 \end{equation}
$n_v$ is the number of valleys, so for Si $n_v=2$.  Equations (1)
and (2) are to be integrated upward in $y$, starting from initial
bare values for the dependent variables at $y = y(0)$ for which
$T_{el}/T$ is of order unity. Increasing $y$ corresponds to
integrating out shorter wavelength and higher energy excitations.
This amounts to increasing the length scale and decreasing the
temperature scale.
 Our Eqs.\ (1) -- (3) are essentially the same as
Eqs.\ (1) and (2) of Ref.\ \onlinecite{Punn}.  The notational
difference is that we use a running variable $y=-\ln(\lambda)$ as
in Ref.\ \onlinecite{Castellani2}, while Ref.\ \onlinecite{Punn}
uses $\xi=-\ln(k_B T \tau/\hbar)$. Prior to renormalization due to
interacting diffusive modes, we can set $z=1$. After the RG flow
has proceeded to the bifurcation fixed point, we set
$y=\ln[(T_{el}/T)^{1/z}]$ where $z$ is the dynamical critical
exponent of the fixed point.

The first term (unity) on the right hand side of  Eq.\ (\ref{alpha}) contains
no electron-electron interactions and arises from weak localization. On scaling
this term enhances the resistivity.  The other term, in the square brackets, is
due to electron-electron interactions in the singlet and triplet particle-hole
spin states and can be negative when $\gamma_2$ is large enough. When it is
negative this term has the opposite trend of reducing the resistivity upon
scaling. The bare value of $\gamma_2$ is given by $\gamma_2^{bare} = -F_0^a /(1
+ F_0^a )$ where $F_0^a<0$ is the spin-antisymmetric Landau parameter. From
Eq.\ (2), $\gamma_2$ increases with rescaling. With $n_v=2$, the expression for
$\alpha(\gamma_2)$ changes sign when $\gamma_2$ reaches $0.46$.

A change in sign of $\alpha(\gamma_2)$, corresponding to a net delocalizing
effect in zero external magnetic field, has no particular {\it a priori}
implication for magnetoresistance. The magnetoresistance data of Refs.\
\onlinecite{Prus,SimmonsWL} can be consistent with our scaling conclusions if,
for example, the localizing contribution in Eq.\ (\ref{alpha}) has a stronger
magnetic field dependence (at small fields) than the delocalizing contribution
from the electron-electron interactions.  Thus the appearance of a weak
localization precursor signature in magnetoresistance data does not necessarily
imply there will be an eventual turn-up in the zero field resistivity at very
low temperatures.

As we integrate Eq.\ (2), $\gamma_2$ increases from its initial
bare value but diverges at a finite value of $y=y_{max}$ provided
${\cal{R}}$ remains finite at $y=y_{max}$.
\cite{Finkelstein1,Finkelstein2,Castellani} It is easy to confirm
that a solution of Eqs.\ (1) and (2) with finite
${\cal{R}}={\cal{R}}_{min}$ and $\gamma_2$ arbitrarily large is
consistent. In this limit Eq.\ (2) can be written
 \begin{equation}
 d\gamma_2/dy=({\cal{R}}_{min}/2)\gamma_2^2\ .
 \label{dgamma2dy}
 \end{equation}
It follows that $\gamma_2$ diverges at a finite $y_{max}$. To verify the
consistency of a finite ${\cal{R}}_{min}$ when $\gamma_2$ diverges, we can
divide Eq.\ (1) by Eq.\ (2) and rearrange obtaining,
 \begin{equation}
 d{\cal{R}}/{\cal{R}}=-d\gamma_2(6\ln\gamma_2)/\gamma_2^2\ .
 \end{equation}
The  integral of the left hand side is finite so it follows that  ${\cal{R}}$
indeed reaches a finite lower limit ${\cal{R}}_{min}$ as
$\gamma_2\rightarrow\infty$. The rescaling cannot be continued beyond the
singularity at $y=y_{max}$ which means that the zero temperature limit cannot
be reached.

In principle, this singularity in the triplet state scattering
amplitude might signal the onset of a magnetic instability in the
system. On the other hand, the divergence might simply be an
artefact of a low order perturbation expansion.
The ultimate fate of this singularity at very low T is not
completely clear. Castellani {\it et al}\ \cite{Castellani2}
showed that a similar divergence in the energy rescaling function
$Z(y)$ can cause the singular point to shift to extremely low $T$.
The paramagnetic metallic regime would then extend over a wide
temperature. More recently, Kirkpatrick and Belitz \cite{KB} and
Chamon and Mucciolo \cite{CM} have indeed found a solution of the
RG equations corresponding to a disordered ferromagnet. The
question of how low the temperatures would be where such
transitions might occur is open.  In the absence of theoretical
guidance on this point we turn to experiment for information on
the triplet spin state scattering amplitude $\gamma_2$ in the
temperature range of the bifurcation.

References \onlinecite{Shaskin} concluded from an analysis of
magnetoresistance data that there would be a ferromagnetic
instability very near the density of the bifurcation. However,
measurements in Refs.\ \onlinecite{Pudalova,Pudalovb} found an
enhancement of the effective g-factor $g^*$, but no  singularity
for densities down to $r_s=8.4$, a range which includes the
bifurcation.  This is interpreted in terms of the
spin-antisymmetric Landau parameter as $g^*=2/(1+F_0^a)$. $F_0^a$
would be $-1$ at the onset of a ferromagnetic instability.  For
$r_s=8$ Ref.\ \onlinecite{Pudalovb} gives $F_0^a \simeq -0.5$,
corresponding to a  spin susceptibility enhancement factor of $2$.
At temperatures well above the bifurcation the bare values of the
electron-electron interaction amplitudes like $\gamma_2^{bare}$
have negligible diffusion corrections. $\gamma_2^{bare}$ is then
related to $F_0^a$ by $\gamma_2^{bare} = -F_0^a /(1 + F_0^a )$.  A
value $F_0^a = -0.5$ corresponds to $\gamma_2^{bare} =1$.  On the
basis of the direct measurement of $g^*$ we will assume that the
triplet spin state scattering amplitude $\gamma_2$ is finite and
well behaved  throughout the density and temperature range of the
experiments we consider.  With this assumption, an explicit RG
equation is not needed for $\gamma_2$.  Nonmagnetic Fermi liquid
behavior of the system is a sufficient condition for a smooth
$\gamma_2$ but may not be necessary.

\section{Bifurcation  in 2D}

To discuss why Eqs.\ (1) and (2) fail to describe a bifurcation in 2D electron
systems, we first recall the corresponding RG results in $d=2+\epsilon$
dimensions with $\epsilon$ small and positive. Making the same physical
assumptions regarding the interacting diffusive modes and in the same one-loop
approximation, Eq.\ (1) becomes
 \begin{equation}
  d{\cal{R}}/dy = -(\epsilon/2){\cal{R}} + \alpha(\gamma_2) {\cal{R}}^2 \ .
   \label{dRdy}
 \end{equation}
 The first term on the right hand side of Eq.\  (\ref{dRdy}) is the
consequence of the $L^{2-d}$ form factor when converting resistivity to
resistance in a space of $d$ dimensions, and the coefficient $\alpha(\gamma_2)$
in the second term is the same as in Eq.\ (\ref{alpha}).

We identify the bifurcation point  as the point at which
$d{\cal{R}}/dy = 0$.  The zero of Eq.\ (\ref{dRdy}) occurs at the
critical value ${\cal{R}}^* = \epsilon/(2 \alpha)$, provided
$\alpha > 0$, where $\alpha$ is the value of $\alpha(\gamma_2)$
when the bifurcation occurs.   In order for the bifurcation to be
a precursor for a quantum critical point, the temperature must be
a relevant variable. This requires that $\tau^{-1} > 0$ in the
linearized flow equation
 \begin{equation}
 d( {\cal{R}} -{\cal{R}}^*)/dy = \tau^{-1} ( {\cal{R}} -{\cal{R}}^*) \ .
  \label{dRdy1}
  \end{equation}
This procedure for identifying a critical point by a linearized flow equation
is standard.  From Eq.\ (\ref{dRdy}), $\tau^{-1} \epsilon/2 > 0$.  The
condition that $\tau^{-1} > 0$ implies that the resistance has positive
(negative) slope with respect to $T$ in the metallic (insulating) regime.

It is clear that a non-trivial root of $d{\cal{R}}/dy=0$ requires at least two
terms and that for the 2D case Eq.\  (1) as presented (that is, Eq.\
(\ref{dRdy}) with $\epsilon=0$) will not be sufficient.   Attempts have been
made to discuss possible metal-insulator transitions in 2D using Eq.\ (1) by
tuning the parameter $\alpha(\gamma_2)$ to zero.  However, a critical point is
a robust property of the entire $d{\cal{R}}/dy$ and cannot be described by the
properties of only a single term $\alpha(\gamma_2) {\cal{R}}^2$ in a series.
Such a procedure is not stable to the addition of higher order terms.  The
simplest modification of Eq.\ (1) that can show a bifurcation and is consistent
with two-component scaling is
 \begin{equation}
 d{\cal{R}}/dy = \alpha(\gamma_2) {\cal{R}}^2 + \beta(\gamma_2)
 {\cal{R}}^3+\ \dots
  \label{dRdy2}
  \end{equation}
The function $\beta(\gamma_2)$ is not known explicitly but its sign can be
determined by the conditions that ${\cal{R}}^*$ and $\tau^{-1}$ are both
positive.  Linearizing Eq.\ (\ref{dRdy2}) about the zero of its right hand side
gives the linearized flow equation (\ref{dRdy1}) with   ${\cal{R}}^*=
-\alpha/\beta$ and exponent $\tau =  \beta/\alpha^2$. Here $\alpha$ and $\beta$
are the values of $\alpha(\gamma_2)$ and $\beta(\gamma_2)$ when the bifurcation
occurs. ${\cal{R}}^*$ and $\tau^{-1}$ are both positive provided $\alpha<0$ and
$\beta>0$.  There is no bifurcation in 2D  if $\alpha(\gamma_2)$ is always
positive. Integrating Eq.\ (\ref{dRdy1}) starting from an initial ${\cal{R}}_0$
at $y=y_0$ gives two families of curves.  A metallic regime and a bifurcation
can thus be described.

We conclude that the bifurcations in the 2D system and the $2+\epsilon$ system
are controlled by  different fixed points. The fixed point
${\cal{R}}^*=\epsilon/\alpha$ for $2+\epsilon$ requires $\alpha>0$ which is in
the range of weak electron-electron interactions. This fixed point becomes
trivial (${\cal{R}}^*=0$) in the $\epsilon\rightarrow0$ limit and plays no role
in 2D. This allows a new fixed point ${\cal{R}}^* = -\alpha/\beta$ to become
physical in 2D at a scale where electron-electron interactions have become
strong enough to change the sign of $\alpha$.   In each case there is only one
fixed point and the physical picture of interacting diffusive modes is correct
for the determination of the universal contributions.

The strength of the electron-electron interactions in 2D is crucial for
generating a bifurcation. At high densities where the electron-electron
interactions are weak, $\gamma_2^{bare}$ is small and $\alpha(\gamma_2^{bare})$
is positive. As the density is lowered the initial $\gamma_2^{bare}$ increases.
Using the data of Ref.\ \onlinecite{Pudalovb} for $g^*$ in Si, the sign change
in $\alpha(\gamma_2^{bare})$ occurs for a density corresponding to
$r_s\simeq3$. This provides an upper limit to the density at which a
bifurcation can occur in Si. For $r_s<3$ the electron-electron interactions are
too weak.

In order to describe quantitatively the resistivity near the bifurcation where
$R\sim1$, as well as in the insulating regime where $R>>1$,  the sum of the
series implied in Eq.\ (\ref{dRdy2}) must be adequately represented.  If the
series is truncated at an arbitrary finite order a spurious divergence at a
finite $y_{max}$ can occur in the insulating region.  While in the insulating
limit ${\cal{R}}$ and $d{\cal{R}}/dy$ are expected to diverge as
$y\rightarrow\infty$ (that is, at $T\rightarrow0$), the divergence in
$d{\cal{R}}/dy$ must  be sufficiently weak that a spurious divergence in
${\cal{R}}$ at a finite $y_{max}$ does not occur.  A linear power law in
${\cal{R}}$ is the strongest growth of $d{\cal{R}}/dy$ at large ${\cal{R}}$
that permits this (with possible logarithmic corrections).  For simplicity, we
maintain the fixed point structure of the low order terms and introduce a
denominator into Eq.\ (\ref{dRdy2}) to represent the net effect of higher order
terms including the linear growth at large ${\cal{R}}$,
 \begin{equation}
 \frac{d{\cal{R}}}{dy} = \frac{\alpha(\gamma_2) {\cal{R}}^2 + \beta(\gamma_2)
 {\cal{R}}^3}{1+\kappa(\gamma_2){\cal{R}}^2}\  .
  \label{dRdy3}
  \end{equation}
Equation (\ref{dRdy3}) satisfies the minimal conditions of having
a bifurcation with two classes of well defined solutions (metallic
and insulating) depending on the choice of initial conditions at
$y=y(0)$.  It may be regarded as a semi-empirical Pad\'e
approximation to the full series.
 Of course, these minimal conditions do not uniquely determine the functional form
 (see also Section IV.\ B).

We focus in this paper on the metallic regime and the close vicinity of the
separatrix. The known perturbative results are contained explicitly in Eq.
(\ref{dRdy3}) so the solution for $y\rightarrow\infty$ and
${\cal{R}}\rightarrow0$ is exact. The strongly insulating limit
${\cal{R}}\rightarrow\infty$ contains additional logarithmic corrections so
Eq.\ (\ref{dRdy3}) is incomplete in this limit, and a detailed discussion of
the deep insulating regime will be given elsewhere.

Close to the separatrix the functions $\alpha(\gamma_2$), $\beta(\gamma_2)$,
and $\kappa(\gamma_2)$ are taken to be slowly varying and so are replaced by
their constant values $\alpha$, $\beta$ and $\kappa$ for $\gamma_2$ near the
start of the bifurcation. Information on these parameters is given by fitting
to experimental data in the following section. Equation (\ref{dRdy3}) can then
be rewritten in the form
  \begin{equation}
\frac{1}{{\cal{R}}}\frac{d{\cal{R}}}{dy} = \frac{1}{\tau}
\frac{\Delta+\Delta^2}{1+\phi(2\Delta+\Delta^2)}\ ,
  \label{dRdyDelta}
  \end{equation}
where
 \begin{eqnarray}
 {\cal{R}}^*&=&-\alpha/\beta \nonumber \\
 {\cal{R}}/{\cal{R}}^*&=&1+\Delta \nonumber \\
 \tau^{-1}&=&\beta{{\cal{R}}^*}^2/(1+\kappa{{\cal{R}}^*}^2) \nonumber \\
 \phi&=&\kappa{{\cal{R}}^*}^2/(1+\kappa{{\cal{R}}^*}^2)\ .
  \end{eqnarray}
    Linearizing Eq.\ (\ref{dRdyDelta}) in $\Delta$ we recover Eq.\ (\ref{dRdy1}), with the solution
 \begin{equation}
 \ln\{|{\cal{R}}-{\cal{R}}^\star|/|{\cal{R}}(0)-{\cal{R}}^\star|\}=\tau^{-1}(y-y(0))\ .
  \label{lnR}
  \end{equation}

Since $\tau$ describes the rescaling of an inverse length squared, it is
related to the critical exponent $\nu$ of the correlation length by
$\tau=2\nu$.  The temperature is introduced by the thermal length
$\ell_{th}\sim1/T^{1/z}$ giving
 \begin{equation}
 |{\cal{R}}-{\cal{R}}^\star| =
 |{\cal{R}}(0)-{\cal{R}}^\star|\left(T_0/T\right)^{1/z\nu} ,
  \label{R}
  \end{equation}
with the same exponent for both the metallic and the insulating branches. The
prefactor defines a temperature scale $T_0 (\delta) \sim |\delta|^{z\nu} T_0$.
Both of these features agree with the observed scaling and with the
phenomenological scaling of Ref.\ \onlinecite{DAMC}.

 \section{Results}
 \subsection{Vicinity of separatrix}

We now compare our results based on Eq.\ (\ref{dRdyDelta}) with experimental
values of the resistivity $\rho(T)$ close to the separatrix obtained for a
Si-MOSFET taken from Fig.\ 1(b) of Ref.\ \onlinecite{Sarachik} (see also Ref.\
\onlinecite{RMP}). Equation (\ref{dRdyDelta}) can be integrated analytically.
The parameters ${\cal{R}}^\star$, $z$, $\nu$, $\alpha$, $\beta$, and $\kappa$
in ${\cal{R}}(y)$ are chosen to give a best fit to the experimental data.  Not
all these parameters are independent of each other.  The experimental data fix
${\cal{R}}^\star=2.8$.  Since ${\cal{R}}^\star$=$-\alpha/\beta$, we can then
consider $\alpha$ to be fixed with $\beta$ the free variable.  The $z$ and
$\nu=\tau/2$ enter together as a product via
$\tau^{-1}(y-y(0)=\ln(T_0/T)^{1/z\nu}$. The temperature $T_0$ just prior to the
bifurcation was taken to correspond to the temperature scale of elastic
scattering $T_{el}=1.75$ K.  From the definition of $\tau$, we have
$\kappa=\tau\beta - 1/{{\cal{R}}^\star}^2$, which relates $\kappa$ to $\beta$.
There are therefore three independent variables $z$, $\nu=\tau/2$ and $\beta$.

From the combination of electric field scaling and temperature scaling of the
resistivity the dynamical critical exponent $z$ is believed to be in the range
$0.8$ to $1.2$.\cite{RMP}  We have therefore made least squares fits to the
experimental data with $b=z\nu$ and $\beta$ as free parameters for the fixed
values of $z$ in the range $0.8\leq z\leq1.2$.  The optimum values of the
fitting parameters are determined by minimizing the root mean square relative
deviation between theory and experiment
\begin{equation}
 D = \frac{1}{N}\left.\sum_{j=1}^{N}
 \frac{\sqrt{({\cal{R}}^{theory}_j-{\cal{R}}^{expt}_j)^2}}
 {{\cal{R}}^{expt}_j}\right.\ ,
  \label{Rrmsq}
  \end{equation}
where $N$ is the total number of points included in the fit.

We have also examined the sensitivity of the fitted parameters to the
temperature and density range of the fits. This is essential because the values
of the least squares fitted parameters can vary with the range of temperature
and density considered.  Eq.\ (\ref{dRdyDelta}) has been established on the
assumption of constant parameters for sufficiently small $T_0(\delta)/T$. Fits
to data will be valid only if the derived parameters are stable with respect to
reducing the maximum allowed $T_0(\delta)/T$, that is by restricting the
temperature and density range of the fits.

We first carried out fits including all of the data points in Fig.\ 1(b) of
Ref.\ \onlinecite{Sarachik}.  To test sensitivity to the range of fit we
successively restricted the allowed data points by the conditions
$T_0(\delta)/T < 1.0$, $0.5$, and $0.25$.  For $T_0(\delta)/T < 0.5$ and fixed
$z$ the $D$ and the values of the fitted parameters become stable. The
coefficients $\beta$ and $\kappa$ vary slowly with $z$.  However the essential
parameters from the point of view of fitting to a universal scaling form are
constant throughout the range of $z$. These values are $D=0.036\pm 0.0005$,
$b=z\nu=1.09\pm0.005$ and $\phi=0.80\pm0.005$, where the uncertainties reflect
the small variations due to $z$. A constant $b$ and $\phi$ is consistent, since
$(2\phi-1)/b$ is the universal coefficient of the first non-linear correction
$\Delta^2$ in Eq.\ (\ref{dRdyDelta}).

 \begin{figure}[h]
   \includegraphics[width=8.1cm]
    {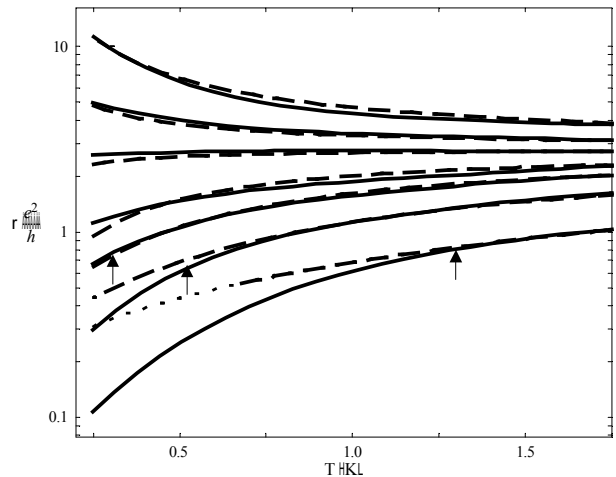}
  \caption{Solid lines: measured resistivity $\rho(T)$ in units of $h/e^2$
from Ref.\ [25] in a Si-MOSFET close to separatrix for electron
densities (from the top) $n=8.6$, $8.8$, $9.0$, $9.3$, $9.5$,
$9.9$, $11.0\times10^{10}$ cm$^{-2}$. Dashed lines: our
calculations using Eq.\ (10).  The values of the parameters are
given in the text.  Dotted lines are in the small $T$ region
$T_0(\delta)/T>0.5$ that is excluded from the fit. Small vertical arrows: see text.
 \label{Fig.1}}
 \end{figure}
Figure 1 compares our results (dashed lines) from Eq.\ (\ref{dRdyDelta}) for
the dimensionless resistivity in the form $\rho e^2/h={\cal{R}}\pi$  with the
experimental measurements (solid lines). The agreement is excellent.  The
dotted lines indicate the data points excluded from the fit by the condition
$T_0(\delta)/T < 0.5$, and the small  arrows show the edge of the corresponding
excluded data range for $T_0(\delta)/T < 0.25$.

\subsection{Exponential form in metallic region}

Previous fits to experimental data have shown that the temperature dependence
of the resistivity is approximately exponential \cite{Pudalove,Pudalovc}. This
result was accounted for by Ref.\ \onlinecite{DAMC} by arguing that the
beta-function of the conductivity was a logarithmic function of conductivity,
even in the metallic range near the separatrix.  This is equivalent to assuming
that the sum of the full series in Eq.\ (\ref{dRdy2}), again with the
coefficients evaluated at the bifurcation, is approximately logarithmic.  Then
Eq.\ (\ref{dRdyDelta}) is replaced by
 \begin{equation}
 \frac{d \ln{\cal{R}}/{\cal{R}}^\star}{dy} \simeq \tau^{-1}\ln{\cal{R}}/{\cal{R}}^\star\  ,
   \label{dlnRdy}
   \end{equation}
  giving
 \begin{equation}
  {\cal{R}}(T)\simeq {\cal{R}}^\star\exp\left\{-C\left[\frac{T_0(\delta)}{T}\right]^{1/z\nu}\right\}
  \label{expR}
  \end{equation}
  with $C$ a positive constant of order unity.

  For the purposes of least squares fitting to data for small $|\Delta|$, it is
  important to note that Eqs.\ (\ref{dRdyDelta}) and (\ref{dlnRdy}) are not
  inconsistent.  Standard procedures for Pad\'e approximations
  give the identity
  \begin{equation}
 \ln(1+\Delta) = \frac{\Delta+\frac{1}{2}\Delta^2}{1+\Delta+\frac{1}{6}\Delta^2}
 +{\cal{O}}(\Delta^5)\ .
  \label{dln1+Delta}
  \end{equation}
  The exact sum of the perturbation series near the bifurcation has the same form
  \begin{equation}
 \frac{d \ln{\cal{R}}/{\cal{R}}^\star}{dy}= \frac{1}{\tau}\frac{\Delta+a \Delta^2}{1+b\Delta+c\Delta^2}
 +{\cal{O}}(\Delta^5)\ ,
  \label{dlnRonR*}
  \end{equation}
with the coefficients being of order unity.  Equation (\ref{dRdyDelta}) and
Eq.\ (\ref{dlnRdy}) with (\ref{dln1+Delta}) can thus be regarded as different
approximations to Eq.\ (\ref{dlnRonR*}) and are therefore equivalent from the
point of view of least squares fitting to data for small $|\Delta|$.  It is
interesting to note that the low order expansion of Eq.\ (\ref{dln1+Delta}),
$\ln(1+\Delta) =\Delta-\frac{1}{2}\Delta^2+{\cal{O}}(\Delta^3)$, gives a value
$\phi=\frac{3}{4}$ which is comparable to our fitted value of $\phi=0.80$.

 \subsection{Deep metallic limit}

Due to its $y$ dependence $\gamma_2$ continues to rescale  as the low
temperature limit of the metallic regime is approached. If we assume that
$\gamma_2$ approaches a finite limiting plateau value $\gamma_2^0$ as
$T\rightarrow0$, then $\alpha(\gamma_2)$ and $\beta(\gamma_2)$ also approach
finite limiting values $\alpha(\gamma_2^0)$ and $\beta(\gamma_2^0)$. Similarly,
again due to the rescaling,  the zero temperature limit $z^0\nu^0$ of the
exponent $z\nu$ is expected to be different from the bare value and also from
the value determined at the bifurcation. In this case for $T$  sufficiently
small we can obtain a consistent solution of Eq.\ (\ref{dRdy2}) with
${\cal{R}}<<1$ so that the perturbation  expansion  is well represented by the
leading term,
 \begin{equation}
 d{\cal{R}}/dy= \alpha(\gamma_2^0){\cal{R}}^2+{\cal{O}}({\cal{R}}^3) \ ,
  \label{dRdy4}
  \end{equation}
with the solution
 \begin{equation}
 {\cal{R}}/{\cal{R}}_0=\left[1-\alpha(\gamma_2^0)(y-y_0)\right]^{-1} \ .
 \label{R2}
 \end{equation}
Using $(y-y_0)=\ln(T_0/T)^{1/z}$ and with the dimensionless
conductivity $g={\cal{R}}^{-1}$, we find
 \begin{equation}
\frac{d\,g\ \ }{d\ln T}= z^0\alpha(\gamma_2^0) \ .
  \label{dgdlnT}
  \end{equation}
  The dominant temperature dependence is then logarithmic  as
observed experimentally.   References \onlinecite{Pudalovc} and
\onlinecite{Pudalovd} express this logarithmic contribution to the
conductivity $G$ as $\Delta G=(e^2/h)C(n)\ln T$.  Using Eq.\
(\ref{dgdlnT}) we identify $C(n)=\alpha(\gamma_2^0)/(\pi z^0)$. As
the density increases $\gamma_2^0$ and $\alpha(\gamma_2^0)$
decrease in magnitude and $\alpha(\gamma_2^0)$ may even change
sign.  This dependence of $C(n)$ on density is in agreement with
that observed experimentally.

\section{Conclusions}

The low order perturbative RG equations of Refs.\
\onlinecite{Finkelstein1,Finkelstein2,Castellani} do not describe a bifurcation
in 2D, but a proper description of a fixed point and a bifurcation in 2D  can
be obtained  when higher order terms in the perturbation expansion for
${\cal{R}}$ are retained (see Eqs.\ (\ref{dRdy2}), (\ref{dRdy3})). The
resulting 2D fixed point exists only for $\alpha(\gamma_2)$ negative and so is
unrelated to the fixed point in dimensions $2+\epsilon$ for which
$\alpha(\gamma_2)$ is positive.

Our results provide a coherent semi-empirical two-component scaling description
of the density and temperature regime near the observed bifurcation and
throughout the metallic regime.  Near the separatrix a least squares fit to
experimental data using Eq.\ (\ref{dRdyDelta}) gives an excellent description
of the observed density and temperature dependence of the resistivity.  A full
discussion for the insulating range will be given elsewhere. The scaling
results apply only to the ``universal'' contributions to the resistivity which
are a generic consequences of the interplay between electron-electron
interactions and disorder.  There are also ``non-universal'' contributions to
the resistivity which will be material-dependent.

The present picture is based on a ${\cal{R}}$ which shows a bifurcation and on
a $\gamma_2$ which scales smoothly with temperature. The bifurcation ``point''
${\cal{R}}^*=-\alpha/\beta$ therefore varies smoothly with temperature and the
separatrix, at which $d{\cal{R}}/dT=0$, is ``tilted'' upwards. A tilted
separatrix can lead in the metallic regime to a turn-up of the resistance at
low temperatures.   This cannot occur in a one-component scaling theory where
the separatrix is flat and the bifurcation point is uniquely determined as a
function of density.

An exact solution of Eq.\ (\ref{dRdy4}) in the very low temperature limit shows
a logarithmic dependence of the conductance on temperature with a coefficient
which is negative and decreases in magnitude as the density increases.   The
low temperature behavior is in agreement with experimental results
\cite{Pudalovc,Pudalovd}. The physical origin of this logarithmic
electron-electron contribution to the conductivity is the same as that observed
at high temperature.  However the numerical value of the low $T$ coefficient
can differ from the bare value at higher temperature.

\begin{acknowledgments}
This work is supported by the Natural Sciences and Engineering Research Council
of Canada and an Australian Research Council Grant.  We thank Alex Hamilton and
Michelle Simmons for very useful discussions.  DJWG thanks Ian Campbell for
hospitality at the Universit\'e de Montpellier II.
\end{acknowledgments}

\end{document}